\documentclass[a4paper,11pt]{article} 
\pdfoutput=1

\usepackage{graphicx,rotating,hyperref,slashed,amsmath,xcolor,amssymb,amsfonts,colortbl,cite,cleveref,booktabs,soul} 
\makeatletter   
\hypersetup{colorlinks,bookmarksopen,bookmarksnumbered,
linkcolor=blus,pdfstartview=FitH,urlcolor=rossos,citecolor=verde}
\allowdisplaybreaks	 
\numberwithin{equation}{section} 
  
\hypersetup{%
    ,urlcolor=black
    ,citecolor=black
    ,linkcolor=black 
    }
\usepackage{mciteplus}

\AtBeginDocument{
  \hypersetup{
    urlcolor=blue,
    citecolor=blue,
    linkcolor=blue,
  }%
}

\newcommand{\be}{\begin{equation}}
	\newcommand{\ee}{\end{equation}}
\newcommand{\bea}{\begin{eqnarray}}
	\newcommand{\eea}{\end{eqnarray}}
\newcommand{\barr}{\begin{array}}
	\newcommand{\earr}{\end{array}}
\newcommand{\ba}{\begin{align}}
	\newcommand{\ea}{\end{align}}

\def\bea{\begin{eqnarray}}
\def\eea{\end{eqnarray}} 

\def\be{\begin{equation}}
\def\ee{\end{equation}} 
\def\beq{\begin{equation}}
\def\eeq{\end{equation}}

\newcommand\ees{\end{eqnarray}}
\newcommand\bees{\begin{eqnarray}}

\def\bea{\begin{eqnarray}}
\def\eea{\end{eqnarray}}

\def\0{{\boldsymbol 0}}

\def\lsim{\mathrel{\rlap{\lower3pt\hbox{\hskip0pt$\sim$}}
   \raise1pt\hbox{$<$}}}         
\def\gsim{\mathrel{\rlap{\lower4pt\hbox{\hskip1pt$\sim$}}
   \raise1pt\hbox{$>$}}}         

 \newcommand{\sfootnote}[1]{}
\definecolor{bluc}{cmyk}{1,1,0,0.1}
\definecolor{rossoCP3}{cmyk}{0,.88,.77,.40}
\definecolor{rosso}{cmyk}{0,1,1,0.4}
\definecolor{rossos}{cmyk}{0,1,1,0.55}
\definecolor{rossoc}{cmyk}{0,1,1,0.2}
\definecolor{verdes}{cmyk}{0.92,0,0.59,0.4}

\hypersetup{colorlinks, bookmarksopen, bookmarksnumbered,
citecolor=verdes, linkcolor=bluc, pdfstartview=FitH, urlcolor=rossos}

\usepackage{multicol}
\usepackage{color}
\definecolor{rosso}{cmyk}{0,1,1,0.4}
\definecolor{rossos}{cmyk}{0,1,1,0.55}
\definecolor{rossoc}{cmyk}{0,1,1,0.2}
\definecolor{blu}{cmyk}{1,1,0,0.3}
\definecolor{blus}{cmyk}{1,1,0,0.6}
\definecolor{bluc}{cmyk}{1,1,0,0.1}
\definecolor{verde}{cmyk}{0.92,0,0.59,0.25}
\definecolor{verdec}{cmyk}{0.92,0,0.59,0.15}
\definecolor{verdes}{cmyk}{0.92,0,0.59,0.4}

\skip\@mpfootins = \skip\footins
\def\circa#1{\,\raise.3ex\hbox{$#1$\kern-.75em\lower1ex\hbox{$\sim$}}\,}

\newfam\rsfsfam
\def\mathscr#1{{\fam\rsfsfam\relax#1}}

\def\circa#1{\,\raise.3ex\hbox{$#1$\kern-.75em\lower1ex\hbox{$\sim$}}\,}
\makeatletter

\def\hhref#1{\href{http://arxiv.org/abs/#1}{arXiv:#1}} 

\newcommand{\doi}[1]{\href{http://dx.doi.org/#1}{[doi]}}

\def\hhref#1{\href{http://arxiv.org/abs/#1}{arXiv:#1}} 
 
\def\art{\@ifnextchar[{\eart}{\oart}}
\def\eart[#1]#2#3#4#5#6{{\rm #2}, {\em #3 \bf #4} {\rm (#6) #5} ({\em #1})}

\def\article{\@ifnextchar[{\earticle}{\oarticle}}
\def\oarticle#1#2#3#4#5#6{{\rm #1}, {\em ``#6''}, {\rm #2 #3 (#5) #4}}
\def\earticle[#1]#2#3#4#5#6#7{{\rm #2}, {\em ``#7''}, {\rm #3 #4 (#6) #5}  [\hhref{#1}]}
\def\hepart[#1]#2{{\rm #2, \em#1}}
\def\heparticle[#1]#2#3{#2, {\em ``#3''} [\hhref{#1}]}

\newcounter{alphaequation}[equation]
\def\thealphaequation{\theequation\hbox to
0.6em{\hfil\alph{alphaequation}\hfil}}
\def\eqnsystem#1{
\def\@eqnnum{{\rm (\thealphaequation)}}
\def\@@eqncr{\let\@tempa\relax \ifcase\@eqcnt \def\@tempa{& & &} \or
  \def\@tempa{& &}\or \def\@tempa{&}\fi\@tempa
  \if@eqnsw\@eqnnum\refstepcounter{alphaequation}\fi
\global\@eqnswtrue\global\@eqcnt=0\cr}
\refstepcounter{equation} \let\@currentlabel\theequation \def\@tempb{#1}
\ifx\@tempb\empty\else\label{#1}\fi
\refstepcounter{alphaequation}
\let\@currentlabel\thealphaequation
\global\@eqnswtrue\global\@eqcnt=0 \tabskip\@centering\let\\=\@eqncr
$$\halign to \displaywidth\bgroup \@eqnsel\hskip\@centering
$\displaystyle\tabskip\z@{##}$&\global\@eqcnt\@ne
\hskip2\arraycolsep\hfil${##}$\hfil& \global\@eqcnt\tw@\hskip2\arraycolsep
$\displaystyle\tabskip\z@{##}$\hfil
\tabskip\@centering&\llap{##}\tabskip\z@\cr}
\def\endeqnsystem{\@@eqncr\egroup$$\global\@ignoretrue} \makeatother

\definecolor{fiorentina}{rgb}{.5,0,.5}

\begin{document}


\setcounter{page}{1} \baselineskip=15.5pt \thispagestyle{empty}

\vspace{0.8cm}
\begin{center}

{\fontsize{19}{28}\selectfont  \sffamily \bfseries {A novel probe of graviton dispersion relations \\ \vskip0.1cm 
at nano-Hertz frequencies}}

\vspace{0.2cm}

\begin{center}
{\fontsize{12}{30}\selectfont  
Bill Atkins$^{a}$ \footnote{\texttt{bill.atkins847.at.gmail.com}}, Ameek Malhotra$^{a}$ \footnote{\texttt{ameek.malhotra.at.swansea.ac.uk}},  Gianmassimo Tasinato$^{a, b}$ \footnote{\texttt{g.tasinato2208.at.gmail.com}}
} 
\end{center}

\begin{center}

\vskip 8pt
\textsl{$^{a}$ Physics Department, Swansea University, SA2 8PP, United Kingdom}\\
\textsl{$^{b}$ Dipartimento di Fisica e Astronomia, Universit\`a di Bologna,\\
 INFN, Sezione di Bologna, I.S. FLAG, viale B. Pichat 6/2, 40127 Bologna,   Italy}
\vskip 7pt

\end{center}

\end{center}

\smallskip

\smallskip
\begin{abstract}
\noindent
We generalise Phinney's `practical theorem' to account for modified graviton dispersion relations motivated by certain
 cosmological scenarios.
 Focusing on specific examples, we show how such modifications can induce characteristic localised distortions -- bumps -- in the frequency profile of the stochastic gravitational wave background emitted from distant  binary sources.   We  concentrate on gravitational waves at nano-Hertz frequencies probed by pulsar timing arrays, and we forecast the capabilities of future experiments to accurately probe parameters controlling modified dispersion relations. Our predictions are based on properties of gravitational waves emitted in the first inspiral phase of the  binary process, and do not 
rely on assumptions of non-linear 
effects occurring during the binary merging
phase.
\end{abstract}

\section{ Introduction}
\label{sec_intro}

The physics of gravitational waves (GW) offers new 
perspectives for testing our understanding of a multitude of physical phenomena and their theoretical interpretations - including providing insights into
gravitational interactions, 
cosmology, and astrophysics (see e.g. \cite{Maggiore:2007ulw,Maggiore:2018sht} for
a pedagogical  treatment of these
subjects). In this work we discuss a novel approach
for  probing the GW dispersion relations
based on a generalization of Phinney's `practical theorem' \cite{Phinney:2001di} to allow for GW  dispersion 
relations different than the case predicted 
by General Relativity. We use propagation
effects of GW travelling cosmological distances.
We remain (for the most part) agnostic to the source of the dispersion relations, but take inspiration by applying this
to the case of modified gravity.

The multimessenger event GW170817 \cite{LIGOScientific:2017vwq,LIGOScientific:2017zic,LIGOScientific:2017ync} set  very stringent bounds on any
deviations of the speed of gravity $c_T$ from the speed of light $c_\gamma$: $|c_T-c_\gamma|\,\le\,3\times 10^{-15}$.
This result
ruled
out many dark energy models based on non-minimal couplings
of extra degrees of freedom with gravity, as shown in \cite{Creminelli:2017sry,Sakstein:2017xjx,Ezquiaga:2017ekz,Baker:2017hug}, building also
on  ideas explored in 
\cite{Lombriser:2015sxa,Bettoni:2016mij,Saltas:2014dha,Nishizawa:2017nef} 
(henceforth we choose units where the speed of light $c_\gamma=1$).
However, it is important to note the GW170817 constraint holds at the Hertz
 frequencies characteristic of LIGO-Virgo-KAGRA (LVK) detections, and
 it is possible that at different scales one finds $c_T\neq 1$,
 due to frequency-dependent modifications
 of the graviton dispersion relations.  In fact, modified gravity models
 of dark energy in the Horndenski class (see e.g. \cite{Joyce:2014kja} for a review) or vector-tensor
 theories (see e.g. \cite{Tasinato:2014eka}) predict a  $c_T\neq1$
 at low energies around spontaneously Lorentz-breaking backgrounds, caused by derivative interactions with the metric~\footnote{Possible exceptions
 can be realized in DHOST scenarios \cite{Langlois:2015cwa,Crisostomi:2016czh,BenAchour:2016fzp}.}. Their Lorentz invariant ultraviolet completions, if  they exist, should
 recover a GW luminal speed at high frequencies  \cite{deRham:2018red}, in particular
 in the LVK band suggesting that the dispersion relations may not yet be directly observable by experimentation. Concrete studies
 based on applications of advanced effective field theory techniques to dark energy set-ups, 
 for example  \cite{deRham:2019ctd}, suggest that a departure from the $c_T= 1$ relation
 should be expected. Moreover, given our ignorance of the real nature of
 dark matter and dark energy, it is important to keep an open mind over scenarios able
 to efficiently scatter GW travelling over  the cosmic medium (as, for example,  solid dark matter \cite{Bucher:1998mh,Zaanen:2021zqs}).
Such scenarios  motivate  modified GW dispersion relations, a topic well explored in the past decades
(see e.g. \cite{Szekeres:1971ss,Madore:1972ww,Madore:1973xy,Press:1979rd}). 

At the present stage, besides the theoretical
aspects of the subject -- currently under interesting development -- it is also important to exploit
forthcoming 
experimental opportunities, and explore phenomenological ways to test modifications 
of standard GW propagation at different  frequencies  with GW experiments. {In particular, the aforementioned scenarios
might  predict rapid changes in the values of $c_T$ as a function of frequency (see e.g. \cite{deRham:2018red})
which motivate the search for methods to test them.}
  From the point of view  of  observations,  apart from
 the ten-Hertz range of LVK measurements, there are existing
 or prospective bounds on $|c_T-c_\gamma|$ at different frequency scales. At  GW  frequencies $10^{-18}\le f/{\rm Hz}\le 10^{-14}$ there are constraints from the CMB \cite{Raveri:2014eea}. At frequencies $ f/{\rm Hz}\simeq 10^{-5}$ there are Hulse-Taylor--type bounds \cite{BeltranJimenez:2015sgd}.  For the $ 10^{-4}\le f/{\rm Hz}\simeq 10^{-2}$ range we will be able to use the red-shift induced waveform dependence  
\cite{LISACosmologyWorkingGroup:2022wjo}
for LISA \cite{Colpi:2024xhw}, as well as its frequency dependence  \cite{Harry:2022zey}.
 For  $10^{-4}\le f/{\rm Hz}\simeq 10^{3}$ there are ideas based on multiband detections \cite{Harry:2022zey}  and   \cite{Baker:2022eiz}. Also the ringing properties of perturbations of black
 hole horizons can, in principle, be used to set constraints
 on $c_T$, see e.g. \cite{Lahoz:2023csk,Mukohyama:2024pqe}.

\smallskip

The list above does not  include nano-Hertz frequencies, which are  interesting to investigate 
   given
 the new possibilities
 offered by  strong hints of detection
 of a stochastic gravitational wave background (SGWB) \cite{NANOGrav:2023gor,Reardon:2023gzh,Xu:2023wog,EPTA:2023fyk}  at those frequency scales (see
 \cite{NANOGrav:2023hvm} for a study within the NANOGrav collaboration to detect new physics with recent Pulsar Timing Array (PTA) data). 
    Recent studies explore modifications of PTA response
 functions \cite{Qin:2020hfy,Liang:2023ary,Cordes:2024oem,Domenech:2024pow} which occur when $c_T\neq 1$, whilst
 various works (see e.g.  \cite{Cannizzaro:2023mgc,Ye:2023xyr,Bernardo:2023mxc,Tasinato:2023zcg,Bernardo:2023zna,Cruz:2024esk}) study further consequences of modified gravity  on PTA data. Here, instead, we consider a  novel effect -- a distortion in the frequency profile of a SGWB produced
 by supermassive black hole  (SMBH) binaries, which is induced by a frequency-dependent
 change in the value of $c_T$ at around the nano-Hertz scales. This effect {exploits   
 propagation effects when GW dispersion relations are modified  and}, if detected,  would allow us
 to experimentally probe a frequency-dependent $c_T$ with PTA experiments. In particular, it allows
 us to measure the case of a $c_T$ varying with frequency, a phenomenon that would be
 difficult to analyze with the alternative methods devised so far. {Moreover, as an important byproduct, it
 can also provide independent information about the redshift  distribution of the sources of SGWB.}
  
  To  develop our understanding of 
  the physics we wish to explore, we  generalise 
Phinney's theorem \cite{Phinney:2001di} to show that the aforementioned  frequency-dependent value of $c_T$  modifies
the frequency profile of the SGWB spectrum -- as discussed in section \ref{gen_phin}. This effect leads to a deformation (a `bump', see Fig \ref{fig:qualityoffit}) on an otherwise power-law   GW spectrum profile. The properties of the bump depend both on the frequency dependence of $c_T$, and on the redshift of the SMBH sources.  Interestingly, the aforementioned
implications  of modified dispersion relations take place during GW emission occurring in the very first inspiral phase of the SMBH merging process, at zeroth order in a post-Newtonian approximation. Hence, it is  not influenced
by screening effects (Vainshtein mechanism, etc) which characterize many modified gravity models, and which due to non-linearities make  a comparison with observations particularly difficult -- see e.g. \cite{Joyce:2014kja} for a review. In Section \ref{sec_fisher}, building on \cite{Ali-Haimoud:2020ozu,Ali-Haimoud:2020iyz}, we perform Fisher forecasts on the prospects of future PTA observations
to set bounds on $c_T$ when monitoring
a large number of pulsars, examining how the result depends
on the modified dispersion relations and on the source properties.
 We conclude in section \ref{sec_conc}.
We set  $c_\gamma\,=\,G_N\,=\,1$.

\section{ A generalization of Phinney's theorem}
\label{gen_phin}

Phinney's theorem relates the energy density in GW with an integral in frequency and redshift of 
quantities associated with GW sources and GW propagation. In fact, in their propagation from source
to detection, GW can experience modified gravity effects that change the standard relation
between GW frequency at emission and detection.  Precisely such cosmological  `redshift' effects, which are well-studied
 in the different context of modifications of GW luminosity distance (see  e.g. \cite{Belgacem:2017ihm,Belgacem:2018lbp}), are at the basis of our 
 generalization of Phinney's theorem in a modified gravity setting. 

To start with, let us develop some basic formulas relating GW 
frequencies  at emission and detection. 
We assume that the GW speed $c_T$ 
depends on the GW momentum, and we restrict the gravitational wave speed to be subluminal, with a dispersion relation taking the form of
\be
\omega(\mathbf{k}) = c_T(\mathbf{k}) \, \mathbf{k}, \quad \quad \mathbf{k} = |\vec{k}|, \quad \quad  c_T(\mathbf{k}) \leq 1.
\ee
To be precise, we must distinguish between the gravitational wave phase velocity, $v_{ph} = \omega(\mathbf{k})/\mathbf{k}$ and the group velocity $v_{gr} = \partial \omega/ \partial \mathbf{k} $. As we wish to remain agnostic to the source of the dispersion relations and model using a simple wavepacket, we presently restrict our analysis to only considering the phase velocity, whereby a complete analysis of the phase and group velocities depend on one's choice of $c_T(\mathbf{k})$ ansatz (further discussion may be found in \cite{Liang:2023ary} and \cite{Ezquiaga_2022}).   See section \ref{sec_intro} for the theoretical motivations supporting this possibility. Now proceeding
along the lines of the arguments developed in \cite{LISACosmologyWorkingGroup:2019mwx,LISACosmologyWorkingGroup:2022wjo} (building on \cite{Maggiore:2007ulw}), we can write the
comoving distance covered by GW along their way to detection as
\be
\label{defrco}
r_{\rm com}^{\rm (GW)}(t)\,=\,\int_0^r\,d \tilde r\,=\,\int_{t_{e}}^t d \tilde t\,\frac{c_T \left[ f (\tilde t) \right]}{a(\tilde t)}
\,,
\ee
where we notice that the comoving distance depends
on the frequency-dependent GW speed $c_T(f)$.  From now on, we  interchange the momentum dependent $c_T({\bf k})$ with a frequency dependent $c_T(f)$, making use of the  implicit relation $k = {2\pi f}/{c_T(f)}$  between frequency and momentum. Let us
 now
consider two wave-crests crossing the same comoving distance, following
   relation \eqref{defrco}. The difference between the times of detection is related
to the difference among time of emission of the wavecrests, by (the suffix $d$  indicates detection position, while $s$ source position)
\be
\Delta t_{d}\,=\,\frac{a(t_d)}{a(t_s)}\,\frac{c_T \left[ f(t_s) \right]}{c_T \left[ f(t_d) \right]}\,\Delta t_{s}\,.
\ee 
This implies that the GW frequencies, proportional to the inverse of the wavecrest $\Delta t$'s, 
satisfy
\be
\label{new_fr_for}
f_s\,=\,\frac{(1+z)}{1-\Delta}\,f_d
\,, \ee
with  $\Delta$ given by
\be
\label{def_delta}
\Delta\,=\,1-\frac{c_T(f_d)}{c_T(f_s)}\,.
\ee
The quantity $\Delta$ controls the frequency-dependent GW speed, and plays
an essential role in our arguments. It  depends on the GW speed $c_T$
at frequencies measured at the position of the source
and of detection. 
Differentiating eq \eqref{new_fr_for}, we find
\be
\frac{d f_s}{f_s}\,=\,\frac{d f_d}{f_d}\,\left( 1-\frac{d\,\ln{(1-\Delta)}}{d\,\ln{(f_d)}}\right)\,.
\ee

These formulas are at the basis of our generalization of Phinney's theorem: we follow 
\cite{Phinney:2001di}, instead using \eqref{new_fr_for} to relate frequencies
at different redshifts.
 The total present-day energy density in GW, as detected by GW experiments, can be expressed as
\be
\label{def_egw}
{\cal E}_{\rm GW}\,=\,\int_0^{\infty}\,\rho_c\,\Omega_{\rm GW}(f_d)\,\frac{d f_d}{f_d}
\,.
\ee
with $\Omega_{\rm GW}$ the  GW energy density for logarithmic frequency interval, divided
by the critical density $\rho_c$.
On the other hand, the quantity
 ${\cal E}_{\rm GW}$ is a sum
of energy densities as radiated at each redshift, taking into account the differential
relation
\be
d {\cal E}_{\rm GW}^{(d)}\,=\,\frac{1-\Delta}{1+z}\,d {\cal E}_{\rm GW}^{(s)}
\,.
\ee
Integrating, we find:

\be
\label{int_rel1}
{\cal E}_{\rm GW}\,=\,\int_0^{\infty} \int_0^{\infty}\,\frac{N(z)}{1+z}\,\left[1-\Delta-\frac{d (1-\Delta)}{d  \ln{(f_d)}}\right]
\,\left(f_s\,\frac{d {\cal E}_{\rm GW}^{(s)}}{d f_s} \right)\,\frac{d f_d}{f_d}\,d z
\,,
\ee

where $N(z)$ is the number of
events for unit comoving volume, occurring between $z$ and $z+d z$. 
Then, equating  relations \eqref{def_egw} and \eqref{int_rel1}, we obtain

\be
\label{phin_gen}
\rho_c\,\Omega_{\rm GW}(f_d)\,=\, \int_0^{\infty}\,d z\,\frac{N(z)}{1+z}\,\left[1-\Delta-\frac{d (1-\Delta)}{d  \ln{(f_d)}}\right]
\,\left(f_s\,\frac{d {\cal E}_{\rm GW}^{(s)}}{d f_s} \right)_{f_s\,=\,\frac{(1+z)}{(1-\Delta)}\,f_d}
\,,
\ee

which represents our generalization of Phinney's theorem. The generalization
includes 
the quantity between parenthesis in eq \eqref{phin_gen}, as well as the evaluation of quantity
at the frequency source $f_s\,=\,{(1+z)\,f_d}/{(1-\Delta)}$ which includes the effects of $\Delta$
as given in eq \eqref{def_delta}.

Before proceeding, let us make a concrete toy example, to explore what we can expect from the previous formula.
Let us assume all sources emit at a common redshift $z_0$, and consist of binaries 
in circular orbits. At the leading (zeroth order) post-Newtonian expansion, the emitted
GW energy reads (see e.g. the textbook \cite{Maggiore:2007ulw})
\be
f_s\,\frac{d {\cal E}_{\rm GW}^{(s)}}{d f_s}\,=\,f_s\,\frac{\pi}{3}\,\frac{{\cal M}^{5/3}}{(\pi f_s)^{1/3}}
\,,
\ee
with   ${\cal M}\,=\,(M_1\,M_2)^{3/5}\,\left( M_1+M_2\right)^{-1/5}$  being
the binary chirp mass. This is the GW energy density emitted during the initial, inspiral
phase of the merging event. We assume
that this formula is valid also in a modified gravity set-up, being derived in a Newtonian approximation. 
Plugging into eq \eqref{phin_gen}, and making use of eq \eqref{new_fr_for}, we find 
\bea
\label{toy_ogw}
\Omega_{\rm GW}(f_d)&=&\frac{8  \left( \pi\, {\cal M}\right)^{5/3}}{9\,H_0^2}\,f_d^{2/3}\,\frac{N(z_0)}{(1+z_0)^{1/3}} \,\left[\frac{
\left(1
-\frac{d \ln \left(1-\Delta \right)}{d \ln f_d}
\right)}{\left(1-\Delta \right)^{1/3}}
\right]
\,.
\eea
We recognize the characteristic, well known $f^{2/3}$ power-law profile of the GW
energy density, although weighted by 
the quantity between squared parenthesis, which depends on the modified GW dispersion
relations  associated with the quantity $\Delta$
of eq \eqref{def_delta}. If measured,
this effect also provides information on the frequency    
dependence profile of $c_T$, a feature that is difficult to extract by other means.
The argument of the squared parenthesis in eq \eqref{toy_ogw} is the frequency $f_d$ at the detector,
which depends on the redshift $z$ of the source (see eq \eqref{new_fr_for}); hence the new part in the squared
parenthesis depends {\it both} on frequency and redshift. Such modified gravity effects, if detected, can then be used
as cosmic ladders {and provide independent
information on the source redshift}.
 To deduce this result, we do not have to
make any assumption on strong gravity effects, nor on the behaviour
of screening mechanisms in specific modified gravity scenarios during the non-linear binary merging process.

While the above formula is obtained for sources
at fixed redshift, it can easily be generalized (at least
formally) to a population of GW sources at different
redshifts. We can write the GW energy
density as
\bea
\label{om_ref}
\Omega_{\rm GW}(f_d)&=&\frac{8  \left( \pi\, {\cal M}\right)^{5/3}}{9\,H_0^2}\,f_d^{2/3}\,
{K}\,{\cal N}_0
\,,
\eea
with  ${\cal N}_0\,=\,\int d z\,N(z)$ and
\bea
K&=& \frac{1}{{\cal N}_0} \int d z\,\frac{1}{\left(1-\Delta(f_d, z)\right)^{1/3}}\left(1
-\frac{d \ln \left(1-\Delta(f_d, z) \right)}{d \ln f_d}
\right)\,\frac{N(z)}{(1+z)^{1/3}} \,.
\label{defbk}
\eea
Details on the source population as a function of redshift affects the factor $K$  in eq \eqref{om_ref},
which depends on the modifications in the GW dispersion relations through the quantity $\Delta$.  
 {Hence, if detected, the effects of modified gravity can  also  probe
the  redshift distribution  of the GW source population.}

\medskip

We proceed by refining and applying the formulas obtained above to 
concrete settings.  For the remainder
of this section, for simplicity, we
focus on sources emitting at a specific redshift $z_0$. (In the next section
we will go beyond this approximation.)
We make a phenomenological
assumption on the behaviour of $c_T$
 as a function of frequency.
 We choose to use the model established in \cite{Harry:2022zey}, namely
\begin{equation}
\label{ansatz1}
c_T(f,\sigma,f_*) = c_0 + (1-c_0) \left(
\frac{1}{2} + \frac{1}{2}\tanh{\Big[\sigma \, \ln\left(\frac{f}{f_*}\right)\Big]} \right)\,.
\end{equation}

\begin{figure}[t!]
\begin{center}
\includegraphics[scale=0.9]{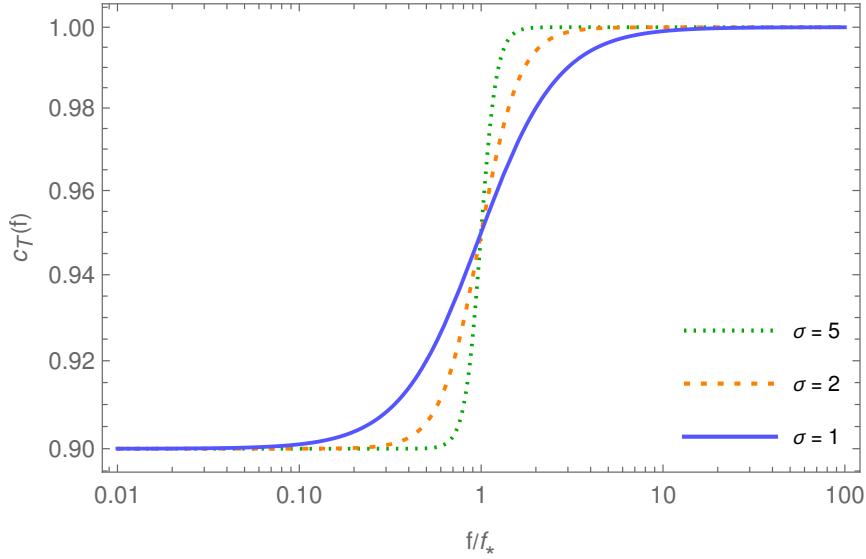}
\end{center}
\caption{\small \it The quantity $c_T$ as a function of $f/f_*$ for $\sigma = 1,\,2,\,5$, 
demonstrating the behaviour of the ansatz \eqref{ansatz1} for $c_0 = 0.9$.}
\label{fig:ctdemo}
\end{figure}

\noindent
\noindent
Schematically, this ansatz for $c_T$ generates a transition at a reference frequency, $f_*$, from $c_0$ to 1, such that $c_T = 1$ at the frequencies detectable by Earth-based interferometers.  If $f_\star$
is sufficiently far from the  frequency band of LVK detections (around 10 Hz), the stringent GW170817 bounds are readily satisfied.   
This choice of dispersion model is not exhaustive, and it is (in part) designed
to describe
the modified gravity scenarios proposed in \cite{deRham:2018red}.
One may also perform the following analysis for several different models with
differing frequency-dependent patterns for $c_T$, as discussed e.g. in \cite{LISACosmologyWorkingGroup:2022wjo}. However our choice of ansatz is particularly easy to manipulate, relying only on three parameters: $\sigma$, controlling the `steepness' of the transition from $c_0$ to $c_\gamma$; $f_*$ controlling the  position of the transition; and $c_0$ being the limit of $c_T$ at low frequencies as shown in Fig \ref{fig:ctdemo}.

 Further to this, one may easily expand upon this ansatz by introducing additional parameters to precisely describe the nature of the transition, or the nature of $c_T$ in the asymptotics. For this work it is sufficient to consider the  simplest case, whereby no additional parameters are included. Taking $\sigma = 1$ and placing eq \eqref{ansatz1} into eq \eqref{def_delta} yields

\begin{equation}
\label{delta1}
\Delta(f_d,\,f_s,\,c_0,f_*) = \frac{c_0 f_*^2(1-c_0)(f_d-f_s)(2c_0 f_* + f_d + f_s)}{(f_d^2 + 2c_0 f_d f_* + c_0 f_*^2)(c_0 f_* + f_s)^2}\,.
\end{equation}

\begin{figure}[t!]
\centering
\includegraphics[width=0.47\textwidth]{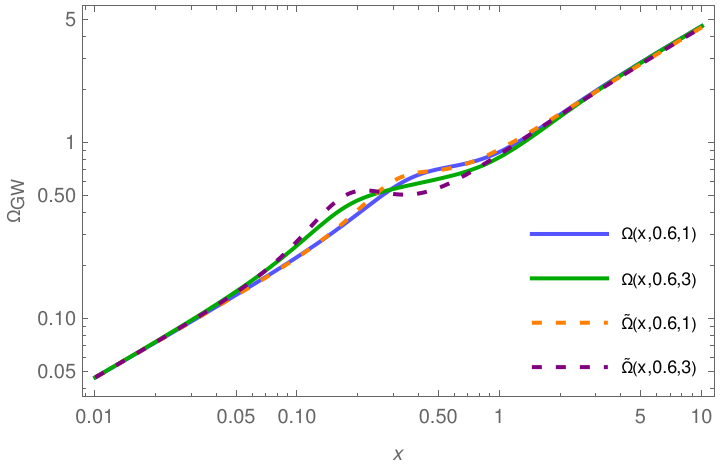} 
\includegraphics[width=0.47\textwidth]{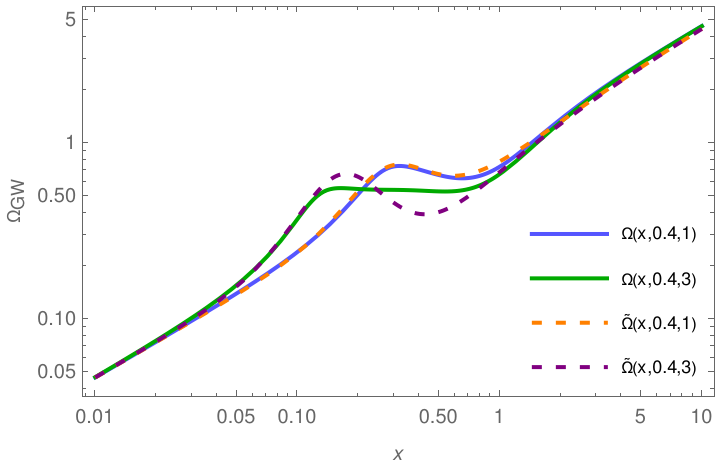} 
\caption{\small \it  Plot of the quantity ${\Omega}_{\rm GW}(x,\,c_0,\,z_0)$ (with  $x = f/f_*$)
for $z_0 \,=\, 1,3$ with $c_0 = 0.6, 0.4$ (respectively left and right panels). We also examine the
the quality of fit of the model in eq \eqref{omegafit}.
 }
\label{fig:qualityoffit}
\end{figure}

Subsequently,
placing eq \eqref{delta1} into eq \eqref{toy_ogw} provides our result for the GW energy
density in the simplest example
of sources emitting at the same redshift $z_0$, 
as in the toy example of eq \eqref{toy_ogw}. We find $\Omega_{ \rm GW}(f/f_*, c_0, z_0)$ produces a deviation from the typical $f^{2/3}$ scaling around the characteristic frequency  $f_*$ characterizing ansatz \eqref{ansatz1},
 with the magnitude and placement of the deviation dependent on the choice of $z_0$ and $c_0$.
 The distorted SGWB frequency profile acquires a characteristic localised `bump' (see Fig  \ref{fig:qualityoffit})
 which may make it distinguishable from other effects such as orbital eccentricities
 or post-Newtonian corrections.  We emphasize that this bump depends
 on the frequency dependence of $c_T$, which can be captured by the modified gravity effects we are presenting here.
 
Increasing the strength of the dispersion (decreasing $c_0$) results in a more sizeable  localized distortion from the $f^{2/3}$  profile, which is magnified at larger redshift, where the magnitude and position of the deviation increases dramatically. We construct a numerical fit for the frequency profile of $\Omega_{ \rm GW}$, in terms of  a combination of Pad\'e approximants to reduce the full expression of $\Omega_{\rm GW}$ to a manageable form. The fit is chosen to asymptote to $f^{{2}/{3}}$, with deviations depending on the  parameters $c_0$ 
characterizing our ansatz \eqref{ansatz1}, and the source
redshift $z_0$.  It reads, defining  $x = f/f_*$ and omitting for simplicity a constant normalization factor (which will be included later), 
\begin{equation}
\label{omegafit}
    \Tilde{\Omega}_{\rm GW}(x,\,c_0,\,z_0) = x^{\frac{2}{3}}\Bigg[ 1+ \frac{30(1-c_0^\frac{1}{3})x^3 \sqrt{z_0}\Big(\frac{3}{\sqrt{1 + c_0} + \frac{5 - 5\sqrt{z_0}}{5+z_0}}-c_0\Big)^3\Big(1.7 + 1.5 c_0 x - \frac{4.5 x}{\sqrt{1 + c_0} + \frac{5 - 5\sqrt{z_0}}{5+z_0}}\Big)}{1+30 c_0 x^5 \Big(\frac{3}{\sqrt{1 + c_0} + \frac{5 - 5\sqrt{z_0}}{5+z_0}}-c_0\Big)^5}\Bigg]\,.
\end{equation}
The previous function reduces to the usual $x^{2/3} \propto f^{2/3}$ profile when $c_0=1$. Notice that
deviations  from the usual profile depend both on $c_0$ and $z_0$, hence they can
be sensitive to redshift . If modified gravity is measured using this method, its effects
can  be used as a
 distance ladder.
To assess the quality of fit we plot in Fig \ref{fig:qualityoffit} a comparison of the fit model and the original $\Omega_{\rm GW}$ with differing choices of $c_0$.
\noindent
Figure \ref{fig:qualityoffit} demonstrates a close correspondence between $\Omega_{\rm GW}$ and the fit in eq \eqref{omegafit} at $z_0 \leq 2$ and $c_0 \geq 0.3$. It also shows  that as $z_0$ increases, the model begins to diverge from the numerical $\Omega_{\rm GW}$. For the purposes of this study we may comfortably restrict to lower $z$, as numerical simulations \cite{Kelley:2016gse} and experimental results \cite{NANOGrav:2023smbhb} show that the peak of the contribution (from astrophysical sources) to the stochastic gravitational wave background is associated with redshifts $z \leq 1$.

Following this theoretical section in which we developed formulas generalizing Phinney's theorem
to modified gravity set-up at fixed redshift,
 we proceed to apply our results to forecast the ability of future Pulsar Timing Array experiments to detect   distortions of the frequency of $\Omega_{\rm GW}$, and further include the redshift dependence of the sources themselves.

\section{ Fisher forecasts}
\label{sec_fisher}

In the previous section we pointed out that a frequency-dependent
GW speed, $c_T(f)$ modifies the predictions of Phinney's theorem.
The effect is associated with propagation properties of GW from
source to detection. 
This effect 
consequently causes
a localized distortion - a `bump' - on the   $f^{2/3}$ frequency profile
characterizing the SGWB emitted by binaries 
in circular orbits, see Fig \ref{fig:qualityoffit}.  In this section, we investigate
whether future experiments can be sensitive to the quantity $c_0$
of eq \eqref{ansatz1} characterizing the deviation of $c_T$ from
luminal speed at small frequencies. We also wish to understand whether
modified gravity effects, if present, can inform us about the nature
of the GW source (redshift, population properties, etc). 
Making use of the Fisher
formalism (see e.g. \cite{Tegmark:1996bz}), 
 we  forecast  the detectability of such effects
 with future pulsar timing arrays (PTA) observations. Given the
 recent strong hint of detection of a SGWB with PTA \cite{NANOGrav:2023gor,Reardon:2023gzh,Xu:2023wog,EPTA:2023fyk}, this topic
 is very timely. We  assume from now on that the SGWB detected
 by PTA measurements is sourced by supermassive
 black hole (SMBH) binaries in a merging process. We separate
the discussion into two parts, depending on our hypothesis on the distribution
of signal
sources.

  So far, PTA measurements are   unable
 to precisely measure the  details of the slope of the SGWB
 frequency profile. In the future, measuring more pulsars and/or
 reducing  the corresponding  noise sources, the situation
 should improve. To address this topic, we adopt an 
 approach  motivated by  \cite{Ali-Haimoud:2020ozu,Ali-Haimoud:2020iyz},
 and we consider an idealized situation of $N_{psr}$ monitored
 pulsars (with $N_{psr}$ a large number)
 isotropically located in the sky, all with the same noise properties.
 This is certainly
 an idealised situation, but in first approximation it mimics
 what we can achieve in the forthcoming SKA era \cite{Janssen:2014dka,Weltman:2018zrl},
 when we will be able to monitor hundreds of millisecond pulsars.

Instead of $\Omega_{\rm GW}$, we focus on the GW intensity ${\cal I}$ as
 a quantity more commonly studied in PTA data analysis. We assume
 the fit $\tilde \Omega_{\rm GW}$ in eq \eqref{omegafit} as a theoretical template
 for the GW energy density, including the effects of modified GW dispersion relations. The GW intensity
 then reads
 \be
 \label{def_int}
{\cal I} \,=\frac{3 H_0^2}{4 \pi^2 f^3}\, \Omega_{0}\, \Tilde{\Omega}_{\rm GW}\,,
\ee
where the overall constant factor $\Omega_{0}$ captures the overall normalization omitted in the fitting
template \eqref{omegafit} for the SGWB frequency profile. 
In measuring the intensity of the GW, we can learn about the amplitude and frequency
dependence of the SGWB, but also set constraints on the quantities $c_0$ and
$z$ characterizing the theoretical fit $\Tilde{\Omega}_{\rm GW}$
of eq \eqref{omegafit}. These are, in fact, the quantities we are interested in and seek to
constrain with future data sets. From now on, we set $\Omega_{0} = 8.1 \times 10^{-9}$  (such that the spectrum aligns with the amplitude measured by NANOGrav in \cite{NANOGrav:2023gor}), and we choose a reference frequency $f_* = 1 \times 10^{-8}$ (to be in line with the order of magnitude of the typical frequencies used for PTA experiments). 

\subsection*{Sources at common redshift $z$}

\begin{figure}[t!]
\centering
\includegraphics[width=0.5\textwidth]{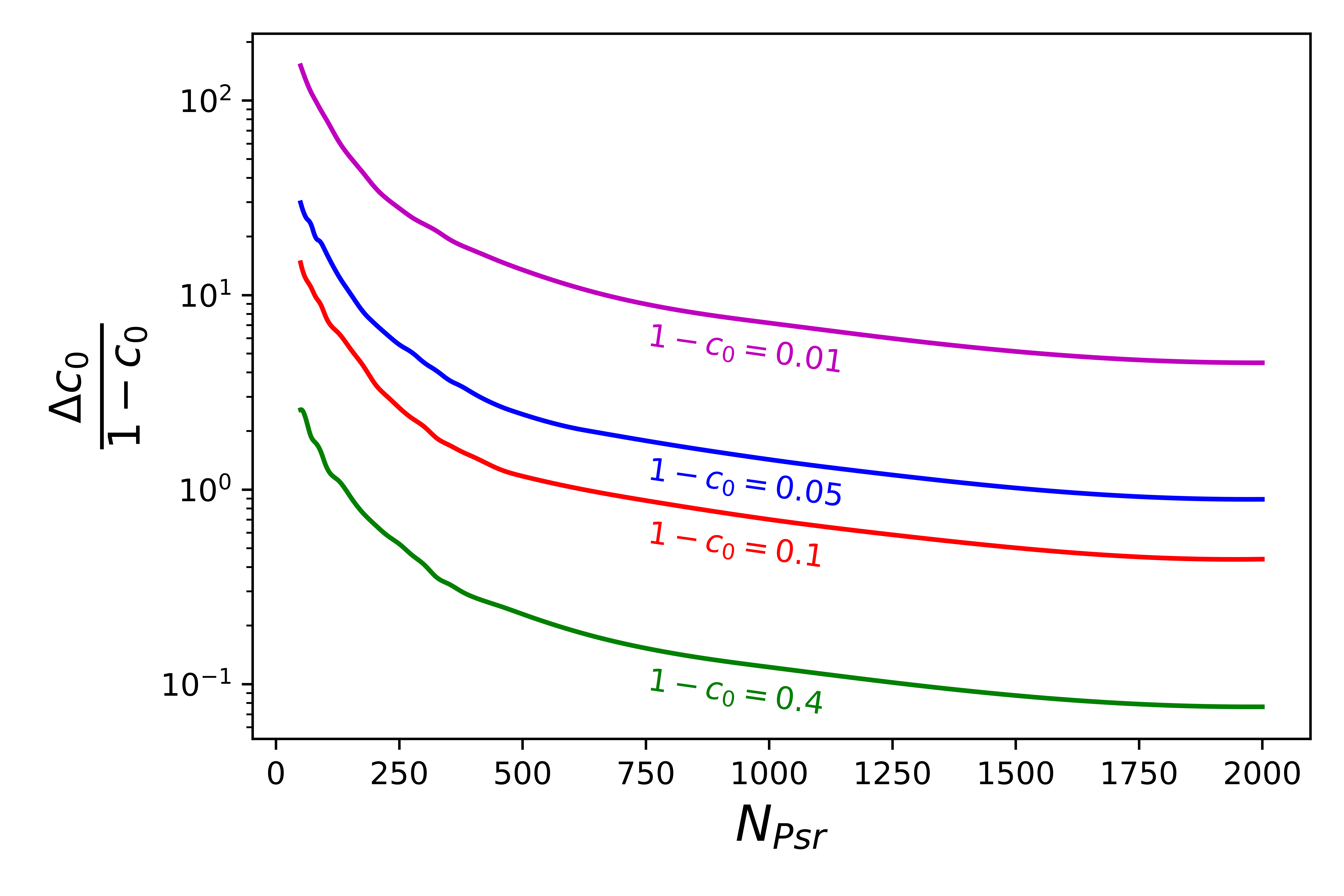} 
\includegraphics[width=0.45\textwidth]{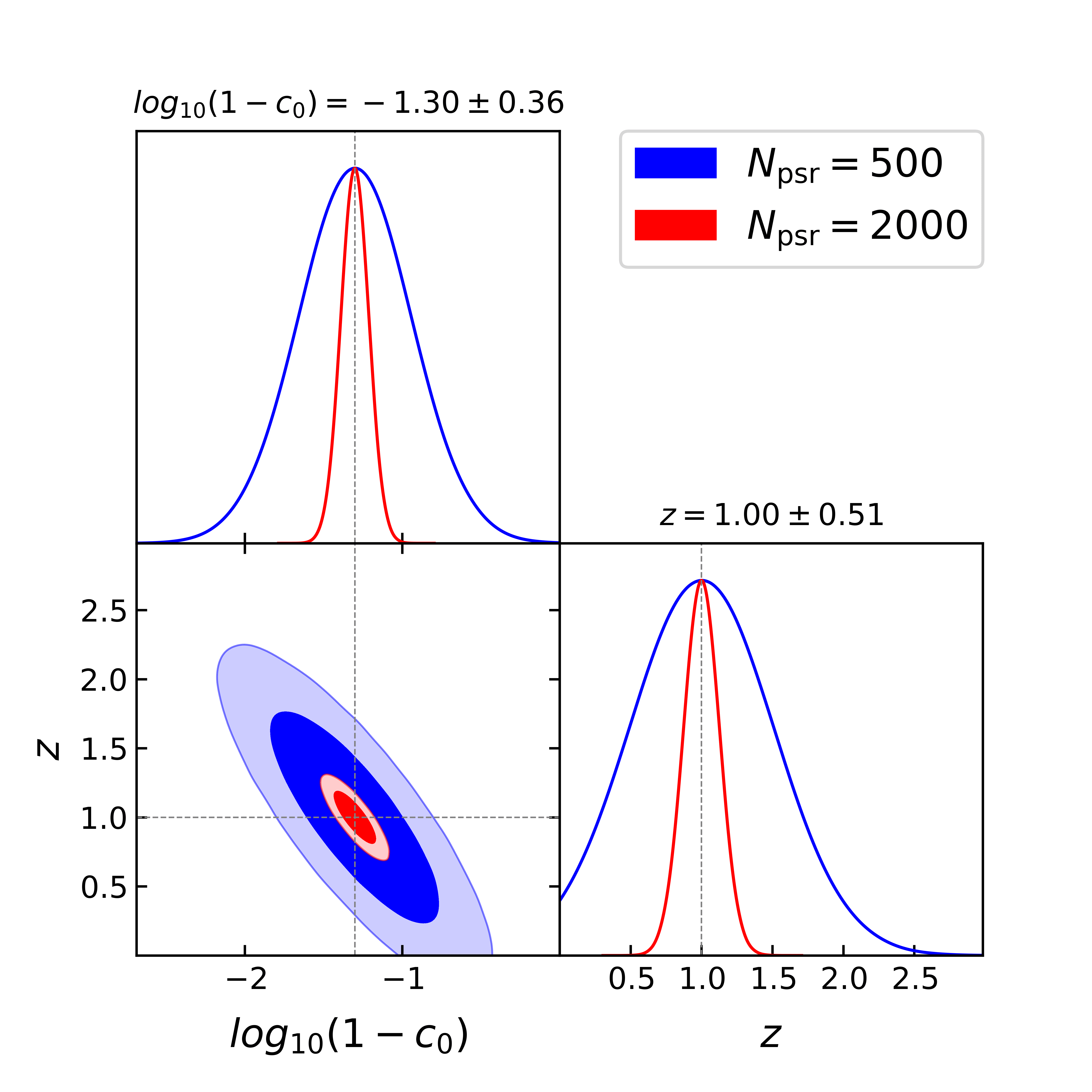} 
\caption{\small  \it {\bf Left panel:} The relative error in $c_0$ as a function of $N_{psr}$ for $1 - c_0 = 10^{-1}$ to $1 - c_0 = 10^{-2}$ marginalized over redshift. {\bf Right panel:} The relative error ellipses for $\log_{10}(1-c_0)$ against $z$ for $N_{psr} = 500, 2000$ and $1-c_0 = 5 \times 10^{-2}$ with all sources located at the benchmark value $z = 1$.  The quoted errors refer to the the $N_{psr} = 500$ case.}
\label{fig:figure3}
\end{figure}

To start our analysis, we make the initial simplification by assuming all SMBH sources occur at a fixed
redshift $z$. 
PTA experiments measure time-delays on millisecond pulsar periods. To detect GW we correlate
measurements between pulsar pairs, each pulsar in the pair denoted with ($a$, $b$). Correspondingly,
we build a Gaussian log-likelihood in terms of the GW intensity, evaluated at a fixed frequency $f$
within a band $\Delta f$:
\begin{equation}
    \mathcal{L}_f = -\log(L) = \frac{1}{2}(d-\mu)\,C^{-1}\,(d-\mu)^T\,,
\end{equation}
where $d$ corresponds to the vector of measured data,  $C_{ij}$ is the  covariance matrix, and $\mu$ represents the vector of length $N_{\rm pair} = N_{psr}(N_{psr}-1)/2$, containing the cross-power spectrum of the gravitational-wave induced time residuals for distinct pulsar pair.  We  denote the latter with $\mathcal{R}_{ab}^{\rm GW}$.
To be explicit, each element of $\mu$ contains $\mathcal{R}_{ab}^{\rm GW}$ plus noise, however as we shall soon search for the expectation values of $\mathcal{R}_{ab}$, we anticipate the noise contributions average to zero (more on this later). 
We assume that the intensity ${\cal I}$ is isotropic
and independent from the GW direction.
 The quantity $\mathcal{R}_{ab}^{\rm GW}$, at a given frequency, results:
\begin{equation}
\mathcal{R}_{ab}^{GW} =
 \frac{\mathcal{I} \cdot \mathcal{H}(\nu)}{(4\pi f)^2}\,,
\end{equation}
with $\nu =\hat{a} \cdot \hat{b}$ being the angle between the pulsar directions (with respect
to the Earth) 
in a given pulsar pair.
 We take for the function $ \mathcal{H}(\nu)$
 the Hellings-Downs formula \cite{Hellings:1983fr}
\begin{equation}
\mathcal{H}(\nu) = \frac{3+\nu}{3} + 2(1-\nu)\ln \Big(\frac{1-\nu}{2}\Big)\,.
\end{equation}
{
In fact, we might expect corrections to this formula for modified dispersion relations, which have
been computed in the limit of frequency-independent, constant $c_T$ (see e.g. \cite{Qin:2020hfy,Liang:2023ary,Cordes:2024oem,Domenech:2024pow}). But we
take the Hellings-Downs result as first approximation in our set-up where $c_T$ has a pronounced 
 frequency dependence  in the PTA band. If the induced effects from modified gravity corrections are small, this remains a reasonable first order approximation; however, it is also important to note that if corrections are prominent at characteristic frequencies, they may help break the degeneracies similarly to those seen in Fig \ref{fig:figure3}. }

We now proceed to label each  pair of distinct pulsars   with capital indices, $I = (a,b)$, as 
in \cite{Ali-Haimoud:2020ozu}. As mentioned above, we make the hypothesis that all pulsars have identical noise properties. {We assume a weak-signal limit where the background signal is subdominant to the intrinsic common pulsar noise, which we define as $\sigma_p$; namely\footnote{We note that the strength of the signal observed by various PTA collaborations means that currently PTAs operate in the intermediate, rather than the weak signal regime. To keep our Fisher analysis simple, we have deliberately chosen noise parameters such that the weak-signal limit applies for our hypothetical future PTA. A more realistic analysis would use the lower noise levels of current PTAs  or the expected noise levels of future PTAs and account for the full covariance matrix. Lowering the noise levels will lead to smaller error bars on the parameters of interest, but only up to a point. Once we reach the intermediate/strong signal limit, the signal variance takes over as the limiting factor in the measurement of the parameters of interest, preventing us from measuring these parameters with arbitrary precision.} 
\begin{equation}
\label{intocheck}
    \mathcal{I} \ll (4\pi f)^2 \sigma_p^2\,.
\end{equation}
}
We select for our analysis a time of observation
\be
T_{\rm obs}\,=\,15\,{\rm years}\,,
\ee
as in the current NANOGrav data set. 
 We are then able to work with a diagonal covariance matrix of the form \cite{Ali-Haimoud:2020ozu}
\begin{equation}
    C_{IJ} \approx \frac{\sigma_a^2 \sigma_b^2}{2 T_{\rm obs} \,\Delta f}
\,.
\end{equation}
 We then generate our noise curve from the \verb|Hasasia| package \cite{Hazboun:2019nqt} with $N_{psr}$ uniformly generated pulsars, choosing (for the white noise parameters) a cadence  $T_{\rm cad}=year/15$, and selecting \mbox{$\Delta t_{\rm rms} = 100\,\rm ns$} to be the rms error of the timing residuals. For the red noise parameters, we select $A_{\rm RN} = 2 \times 10^{-15}$, and $\alpha_{\rm RN} = -2/3$ in line with \cite{cruz2024measuringkinematicanisotropiespulsar} such that we remain within the weak-signal limit.
  After these considerations, we continue our discussion.
 We  choose the size of the frequency bands $\Delta f = 1/T_{\rm obs}$. 
  Hence, our covariance matrix  becomes proportional to the identity
\begin{equation}
C_{IJ}^{-1} \approx \frac{2}{\sigma_p^4} \delta_{IJ}\,,
\end{equation}
and the log-likelihood reduces to a  simple expression at a given frequency
\begin{equation}
\mathcal{L}_f = \frac{2}{(4\pi f \sigma_p^2)^2}\delta_{IJ}
\,
\mathcal{H(\nu)}
\,
\mathcal{I}(f/f_\star,c_0,z)\,.
\end{equation}
The full likelihood requires a sum over all frequency bands: $\mathcal{L} = \sum_{band(f)}\,\mathcal{L}_f $. We choose   these bands ranging from $1/T_{\rm obs}$ to $20/T_{\rm obs}$.
 The Fisher matrix is obtained by computing second derivatives of the log likelihood over the components $\theta_\alpha$ of
the  vector
$\vec \theta$
of  parameters of interest, and taking the expectation value:

\begin{equation}
\label{fisher1}
F_{\alpha \beta} = \left \langle \frac{\partial^2 \mathcal{L}}{\partial \theta_\alpha \partial \theta_\beta}\right \rangle
\,.
\end{equation}
We restrict the vector of parameters to constrain to the quantities
  $c_0$, and $z$ (recall that we are assuming that all sources
  are located at the same redshift). The Fisher matrix evaluates to 

\begin{equation}
\begin{split}
F_{\alpha \beta} & =  \frac{1}{2} \sum_{band(f/f_*)} {\rm Tr} \Big(C_{IJ}^{-1} \,\partial_\alpha C_{IJ} \, C_{IJ}^{-1} \,\partial_\beta C_{IJ} + C_{IJ}^{-1}(\partial_\alpha \mu \, \partial_\beta \mu^T + \partial_\beta \mu \, \partial_\alpha \mu^T)\Big) \,,\\
& = \frac{1}{\sigma_p^4} \sum_{band(f/f_*)}\big(\partial_\alpha \mu \, \partial_\beta \mu^T + \partial_\beta \mu \, \partial_\alpha \mu^T \big)\,, \\
& = \frac{1}{(4 \pi f)^2 \sigma_p^4} \sum_{band(f/f_*)}
    \begin{pmatrix}
        2 \frac{\partial \mathcal{R}_{ab}}{\partial {c_0}} \Big(\frac{\partial \mathcal{R}_{ab}}{\partial {c_0}}\Big)^T  & \frac{\partial \mathcal{R}_{ab}}{\partial {c_0}} \Big(\frac{\partial \mathcal{R}_{ab}}{\partial {z}}\Big)^T + \frac{\partial \mathcal{R}_{ab}}{\partial {z}} \Big(\frac{\partial \mathcal{R}_{ab}}{\partial {c_0}}\Big)^T\\
         \frac{\partial \mathcal{R}_{ab}}{\partial {z}} \Big(\frac{\partial \mathcal{R}_{ab}}{\partial {c_0}}\Big)^T + \frac{\partial \mathcal{R}_{ab}}{\partial {c_0}} \Big(\frac{\partial \mathcal{R}_{ab}}{\partial {z}}\Big)^T & 2 \frac{\partial \mathcal{R}_{ab}}{\partial {z}} \Big(\frac{\partial \mathcal{R}_{ab}}{\partial {z}}\Big)^T \\
        \end{pmatrix} \,.\\
\end{split}
\end{equation}

\noindent
 The marginal error in the parameter $\theta_\alpha$ we wish to constrain is given by
\begin{equation}
\Delta \theta_\alpha \geq \sqrt{(F)^{-1}_{\alpha \alpha}}\,.
\end{equation}
\noindent
From these formulas, focusing
on the quantity $c_0$ we wish to constrain (see eq \eqref{ansatz1}), 
 we determine its  error $\Delta c_0$, as a function of the number of pulsars $N_{psr}$, marginalized
 over the redshift. We represent the result in the left panel of Fig \ref{fig:figure3},  plotting the relative error of the deviation $1-c_0$, (again, marginalised over the redshift). For the generating the plots in the Fisher analysis we make use
of the \verb|GetDist| package~\cite{Lewis:2019xzd}.

The plot  exhibits a linear scaling of the error with the pulsar number $N_{psr}$, such that a deviation of the order (for example) $10^{-1}$ may be discernible within a $10\%$ relative error for $N_{psr}  \sim 1000$.  We may consider
this case more carefully, and explore whether data, besides $c_0$, can also independently constrain the common source redshift $z$ (whose benchmark value is
here $z=1$). We represent error ellipses in the right panel of Fig \ref{fig:figure3}, showing that in fact we can 
obtain a good precision (of order $10-20\%$) in the determination of $z$ with $c_0$, if we monitor
a sufficient number of pulsars. One may find that for $\log_{10}(1-c_0) = -1.3$, a 1-$\sigma$ result may be totally visible at $N_{psr} = 500$, and a 2-$\sigma$ result may be visible for $N_{psr} = 2000$. 
Deviations from
standard dispersion relations, then, can also be used as a ladder to determine the distance from the source. It would
be interesting to further develop this topic, also studying possible ways to resolve degeneracies with $1-c_0$
that are present in the right panel of Fig \ref{fig:figure3}. Intuitively, one possible source of degeneracy for this model may simply occur through the strength of the signal, with sources at lower $z_0$ increasing the magnitude of the bump characteristically associated with the strength of the deviation, $c_0$. Upon further analysis this may be resolvable by closer inspection of the source distribution model over a variety of choices of $z_0$. We then choose to proceed along the lines of improving the source model in the proceeding section.

\subsection*{Refining the model - Integrating over redshift}
\label{refining}

In the analysis
so far we hypothesized that all sources
appear at a common redshift.
We now   relax this assumption  and work with the SMBH population model developed in   \cite{Sesana:2008mz}, as used in \cite{Sato-Polito:2023spo}. See also \cite{Sato-Polito:2024lew} and references therein  for a recent study with
 refinements on SMBH population scenarios. We already
briefly commented  around eq \eqref{defbk} about the mechanism by which our generalization
of Phinney's theorem can accommodate integration over sources
at distinct redshifts. The integration primarily depends on the  dependence
of the pulsar number over redshift. Distortions of the SGWB profile induced by modified dispersion
relations can be used to probe the redshift dependence of the SMBH population. 
Concretely, we  extend our previous analysis, and wish to integrate the fitting formula \eqref{omegafit} over
 redshift. We may then define 
\begin{equation}
\label{newmodeleqn}
{\Omega}_{\rm GW}(x,c_0,z) = \frac{1}{N_0} \int_0^z \Tilde{\Omega}_{\rm GW}(x,c_0,\Tilde{z}) N(\Tilde{z}) \, d\Tilde{z}\,,
\end{equation}
where $N(z)$ is the population at a given redshift, and $N_0=\int dz N(z)$. The number density of sources emitting per redshift, per logarithmic chirp mass is obtained as
\begin{equation}
\frac{dn}{dz \, d\log_{10}\mathcal{M}} = \dot{n}_0 \Big( \frac{\mathcal{M}}{10^7 \mathcal{M}_{\odot} }\Big)^{-\alpha}e^{-\mathcal{M}/ \mathcal{M_*}}(1+z)^\beta e^{-z/z_0}\frac{dt_r}{dz}.
\end{equation}
The quantity $t_r$ represents the coordinate time in the source frame,  the parameters $\alpha$ and $\mathcal{M_*}$ govern the chirp mass distribution, whilst the parameters $z_0$ and $\beta$ govern the redshift distribution (see
\cite{Sesana:2008mz} for more details). The normalized merger rate, $\dot{n}_0$ is fixed such that the overall amplitude matches the profile in eq \ref{def_int} - this may be fixed more rigorously with a treatment such as in \cite{Chen_2019}, however our simple fixing remains within the same bounds. One may then rewrite this in terms of the number of binaries in a spherical shell of thickness $dz$ emitting at a frequency $f_r$ such that

\begin{figure}[t!]
\begin{center}
\includegraphics[scale=0.6]{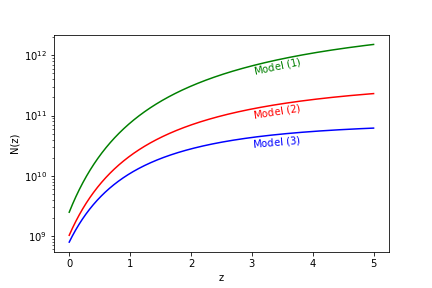}
\end{center}
\caption{\small \it The SMBH binary population models (1)-(3) of Table \ref{table:1}, as a function of $z$ for binaries with chirp mass $\mathcal{M} = 3.2 \times 10^7\mathcal{M}_{\odot}$ emitting at the frequency $f_*$.}
\label{fig:figure7}
\end{figure} 

\begin{equation}
\frac{dn}{dz \, d\log_{10}\mathcal{M}} = \frac{dN}{dz \, d\log_{10} \mathcal{M} \, d\log_{10} f_r} \frac{d \log_{10} f_r}{dt_r} \frac{dt_r}{dz}\frac{dz}{dV_c}\,,
\end{equation}
\\
\noindent
with $V_c$ representing the comoving volume where
\begin{equation}
    \frac{dt_r}{dz}\frac{dz}{dV_c} = \frac{1}{4 \pi(1+z)D_A^2}\,,
\end{equation}
\noindent
and $D_A$ being the luminosity distance \cite{Maggiore:2007ulw}. 
Further assuming the only change in $f_r$ occurs due to energy loss from gravitational radiation yields

\begin{equation}
\frac{d \log_{10}f_r}{dt_r} = \frac{96}{5}\pi^{8/3}\mathcal{M}^{5/3}f_r^{8/3}\,.
\end{equation}

\begin{table}[h!]
\centering
\begin{tabular}{||c c c c c||} 
 \hline
 Model & $\alpha$ & $\mathcal{M}_* [\mathcal{M}_{\odot}]$ & $\beta$ & $z_0$ \\ [0.5ex] 
 \hline\hline
  (1) & 1 & $3.2 \times 10^7 $ & 3 & 3 \\
 (2) & 0.5 & $7.5 \times 10^7 $ & 2.5 & 2.4 \\ 
 (3) & 0 & $1.8 \times 10^8 $ & 2 & 1.8 \\ [1ex] 
 \hline
\end{tabular}
\caption{\it The benchmark population model parameters we consider. }
\label{table:1}
\end{table}

\noindent
Explicitly,  we work with population models (1)-(3) examined in \cite{Sato-Polito:2023spo} (and originally in \cite{Sesana:2008mz}).
  We report the benchmark values
for the corresponding parameters in Table \ref{table:1}.
The population as a function of redshift is then 

\begin{equation}
\label{popmodel}
N(z) = \frac{5 \dot{n}_0}{24 \pi^{5/3}} \int \frac{1}{\mathcal{M}^{8/3}}\Big(\frac{\mathcal{M}}{10^7\mathcal{M}_{\odot}}\Big)^{- \alpha} e^{-\frac{\mathcal{M}}{\mathcal{M}_*}}\, d\mathcal{M}\int \frac{1}{f_r^{11/3}} \, d f_r \int D_A^2 (1+\Tilde{z})^{\beta + 1} e^{-\frac{\Tilde{z}}{z_0}} \, d\Tilde{z}.
\end{equation}

Working such that $\mathcal{M} = 3.2 \times 10^7 \mathcal{M}_{\odot}$ appears as the upper limit of the integration over mass, and again taking the reference frequency $f_* = 1\times10^{-8}$, we can numerically integrate to find a population structure shown in Fig \ref{fig:figure7}. This population model counts the number of contributing sources out to a given redshift, $\tilde{z}$, whose contribution to the background is observed in equation \ref{newmodeleqn}.  For this analysis we assume $H_0 = 70$ km s$^{-1}$ Mpc$^{-1}$, $\Omega_M = 0.3$, $\Omega_\Lambda = 0.7$, and $\Omega_k = 0$.

\begin{figure}[t!]
\centering 
\includegraphics[width=0.32\textwidth]{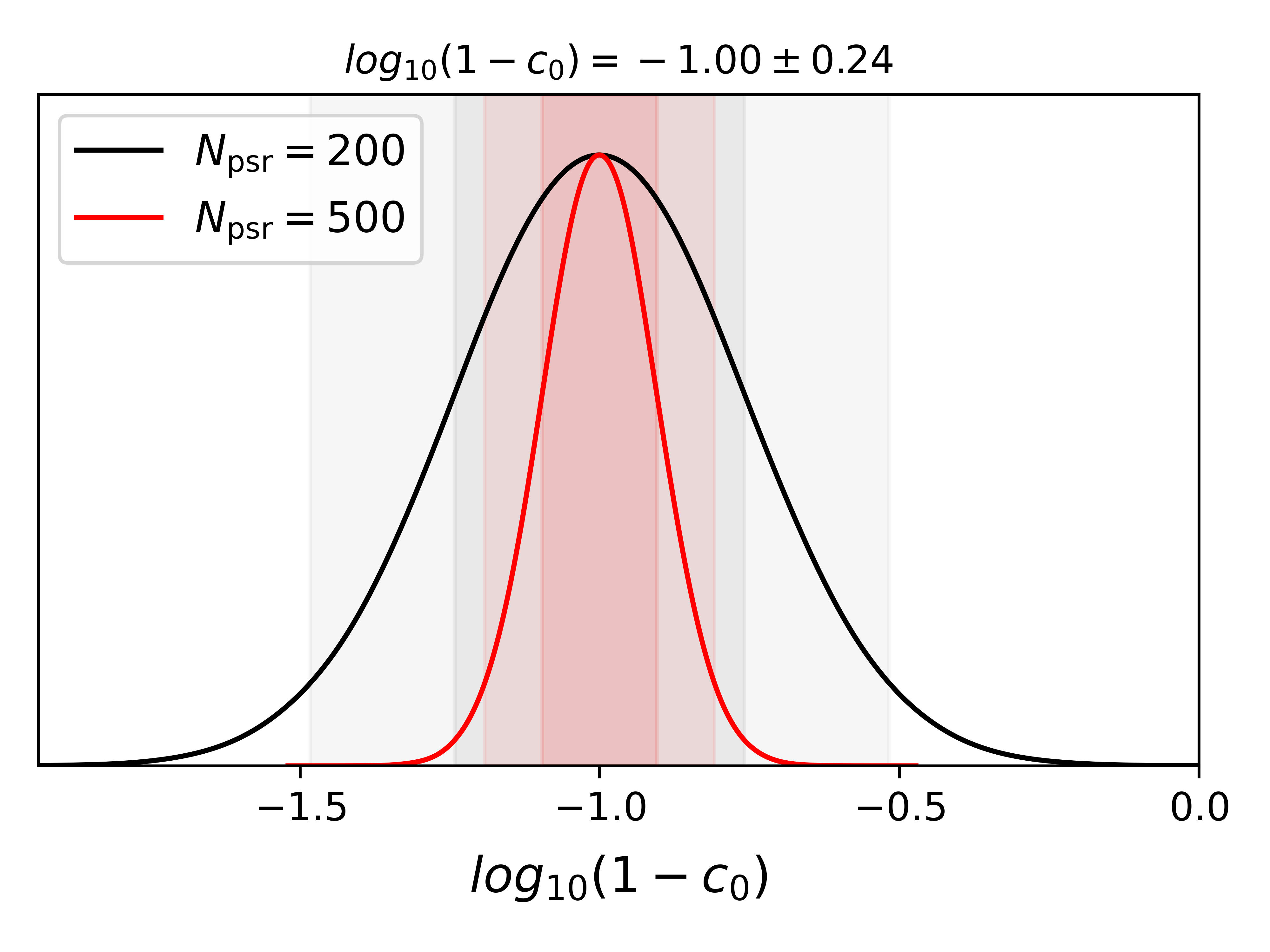} 
\includegraphics[width=0.32\textwidth]{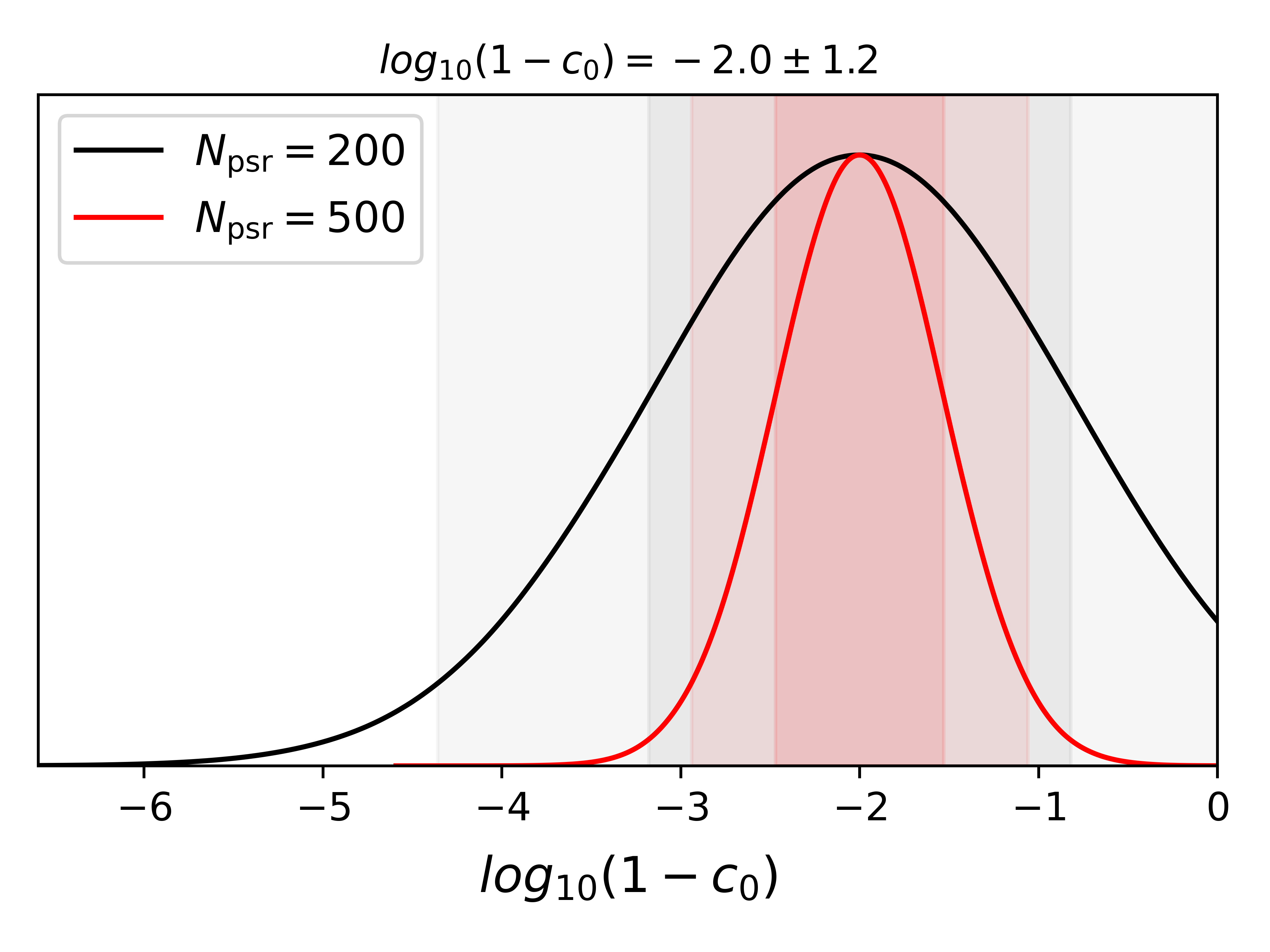}
\includegraphics[width=0.32\textwidth]{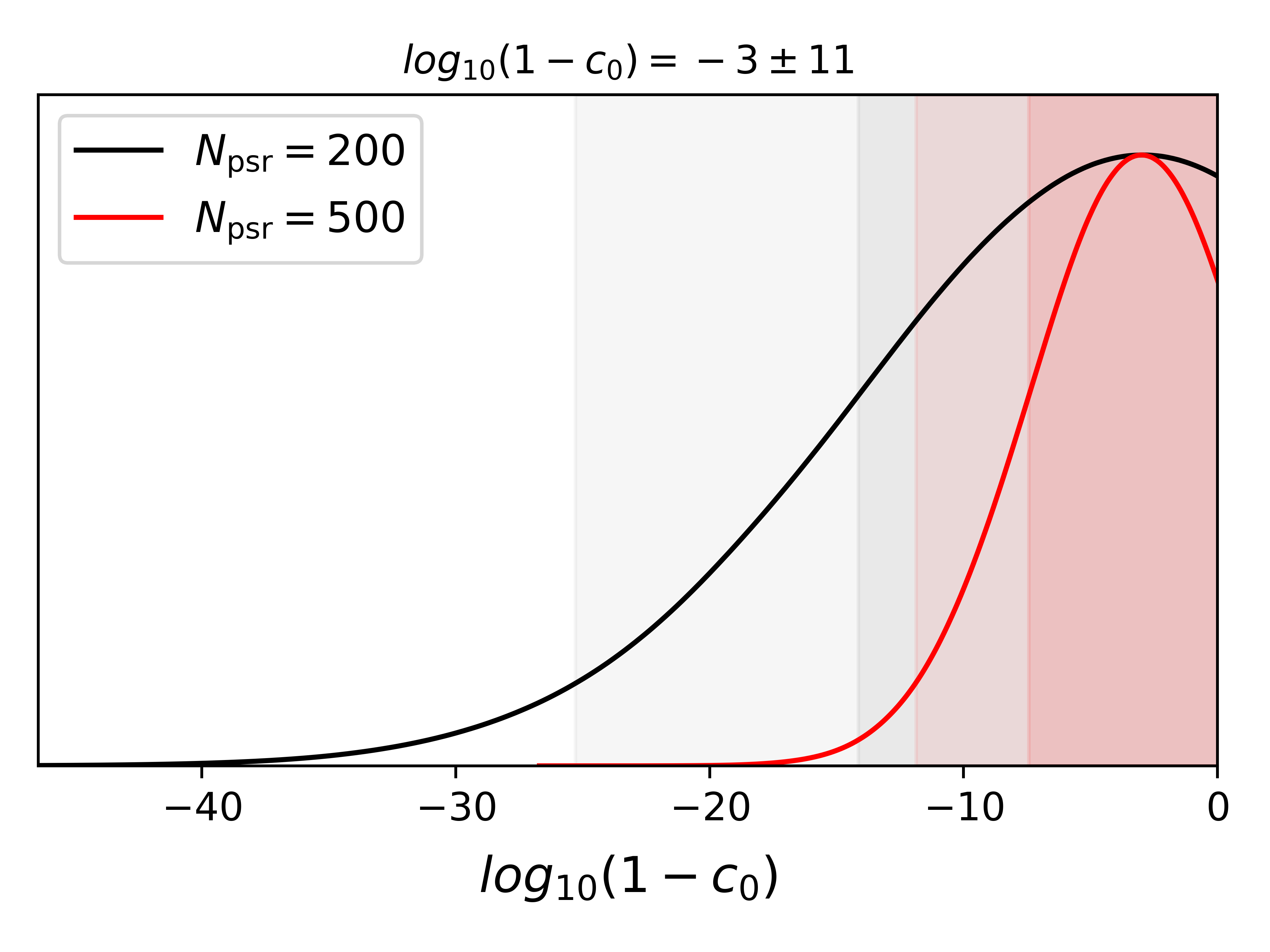}
\includegraphics[width=0.32\textwidth]{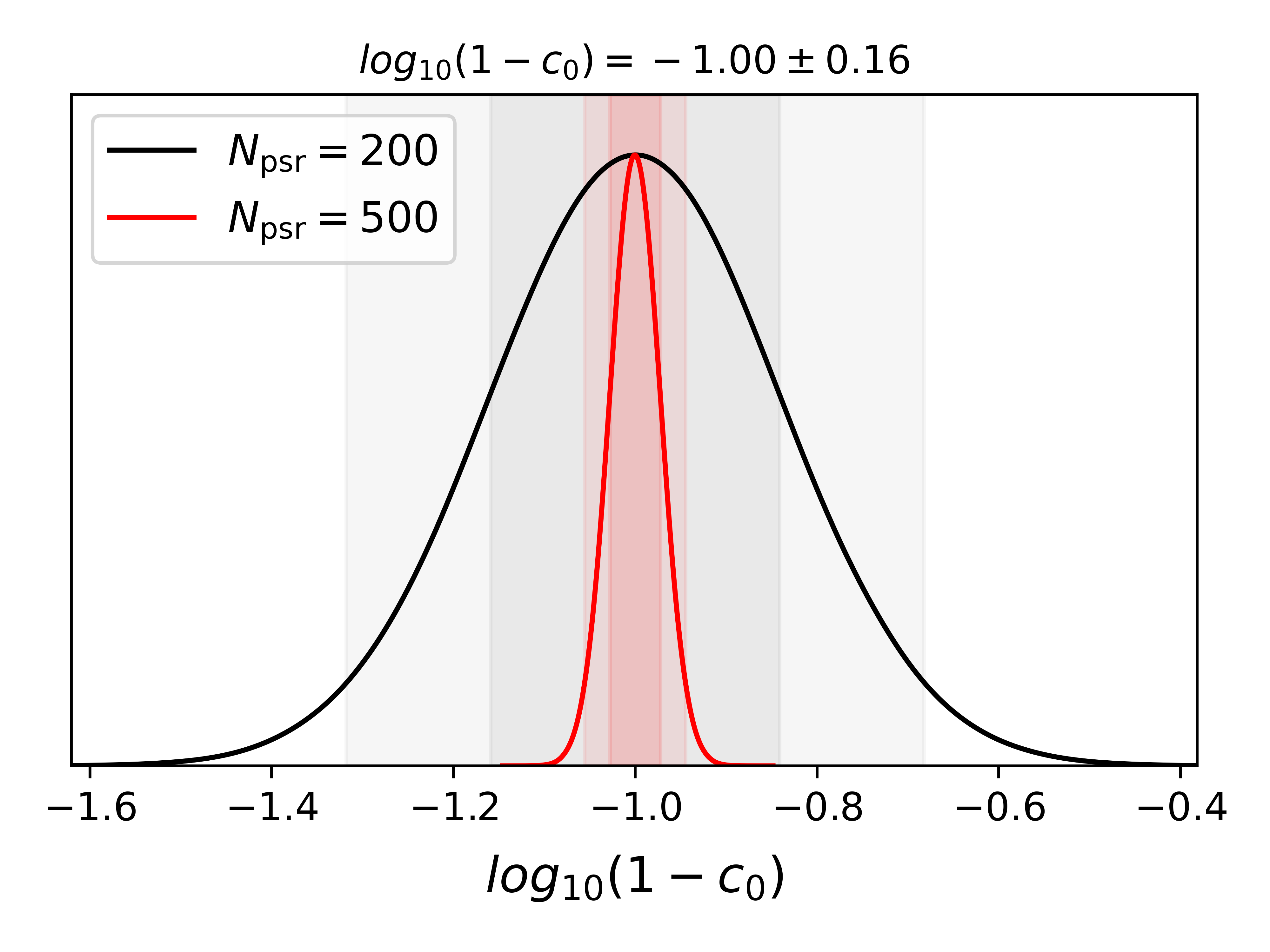} 
\includegraphics[width=0.32\textwidth]{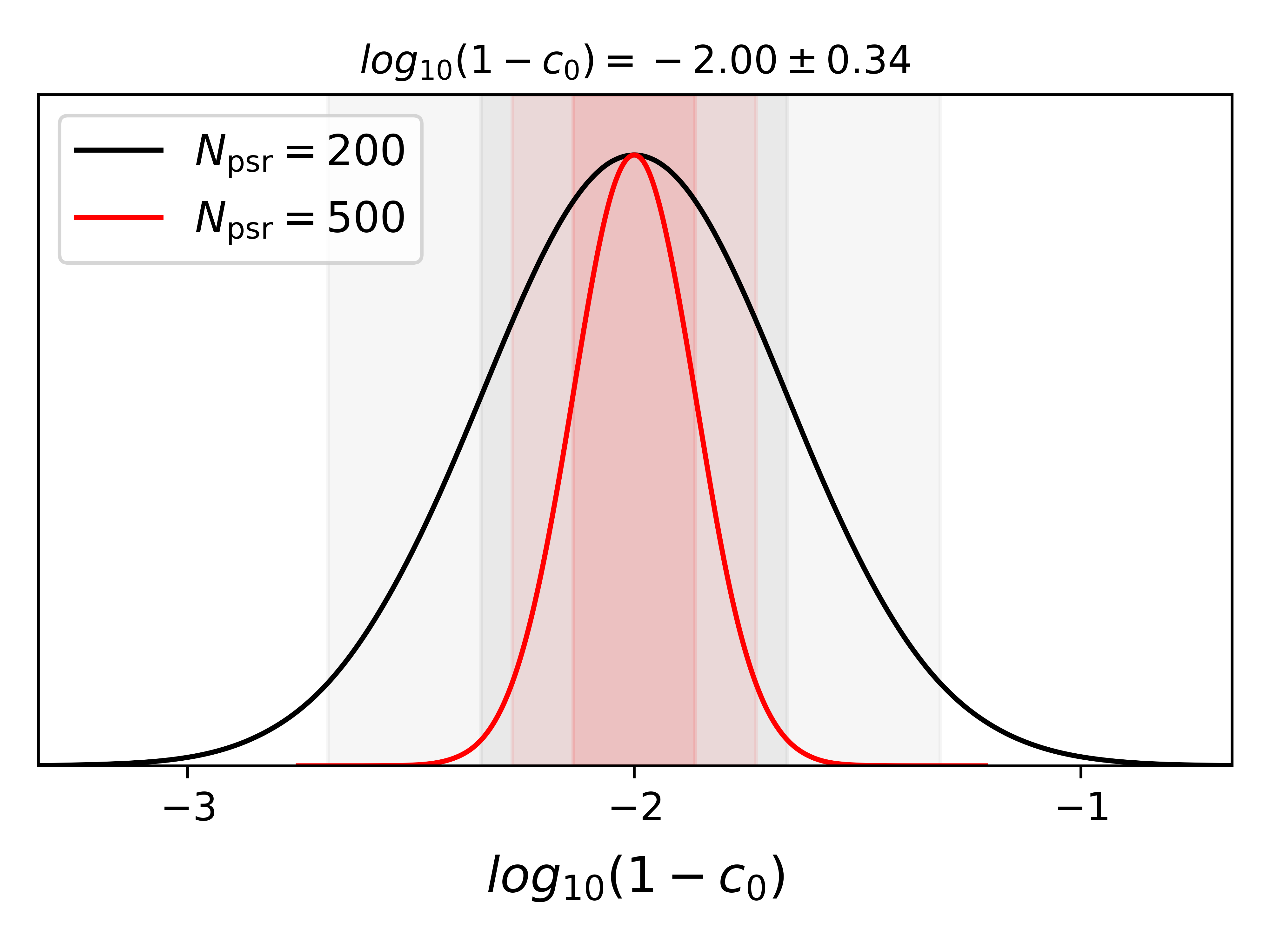}
\includegraphics[width=0.32\textwidth]{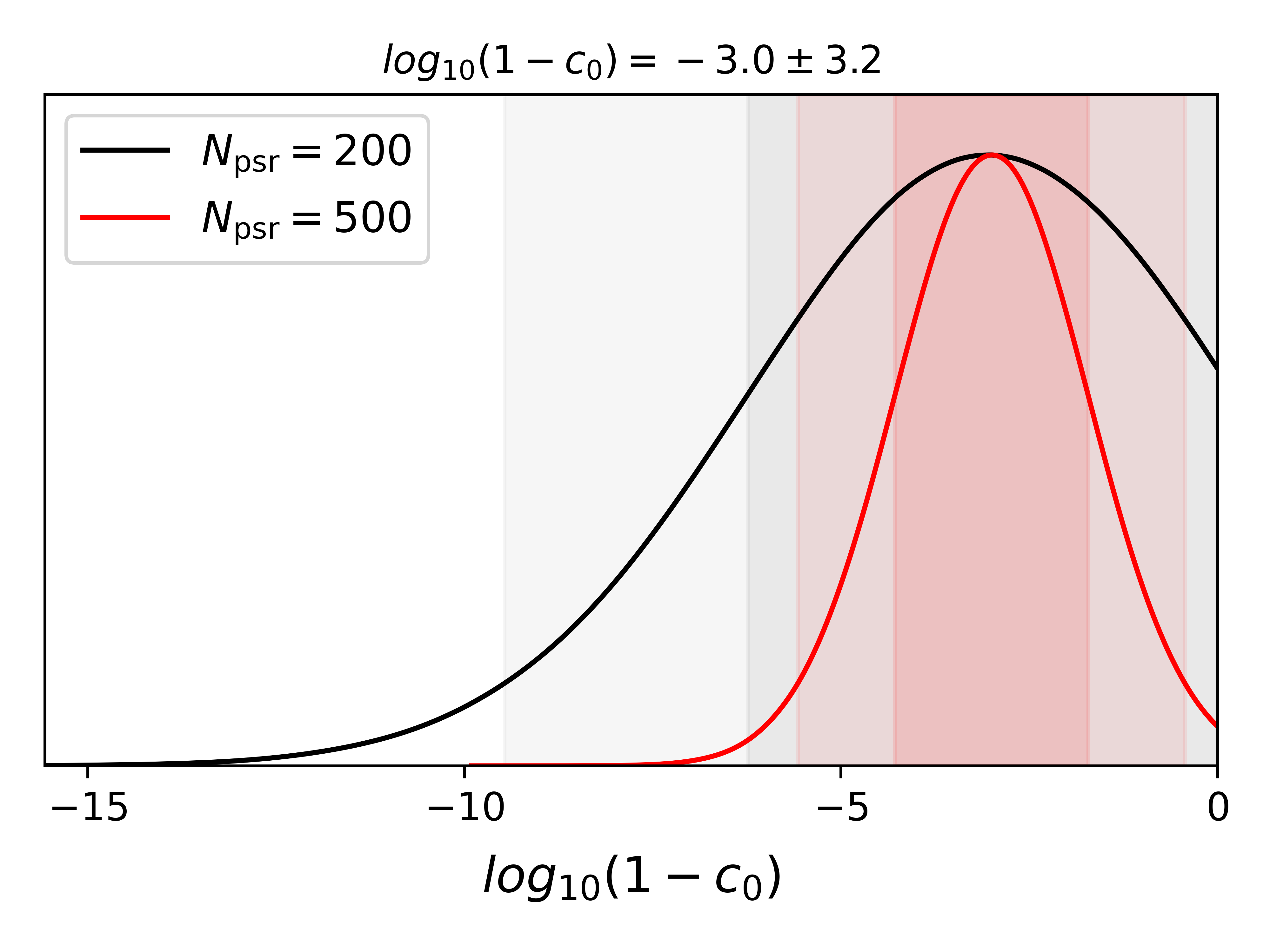}
\caption{\small  \it
The relative error bounds for $\log_{10}(1-c_0) = -1, -2, -3$ benchmark values,    integrating
over redshift eq \eqref{popmodel} to $z = 1$. {\bf First row:}   population model $(1)$. {\bf Second row:}   population model $(3)$.
 We select $N_{psr}\,=\,200$ and $N_{psr}\,=\,500$. In all plots, the darker bands represent a 1-$\sigma$ deviation, and the lighter bands represent a 2-$\sigma$ deviation. The quoted errors refer to the $N_{psr} = 200$ case with all errors tabulated in Table \ref{table:2}.
}
\label{fig:figure8}
\end{figure} 

\begin{table}[h!]
\centering
\begin{tabular}{||c |c |c |c |c||} 
 \hline
$\log_{10}(1-c_0)$ & \multicolumn{2}{|c|}{$\Delta \log_{10}(1-c_0), \ N_{psr} = 200$}  & \multicolumn{2}{c||}{$ \Delta \log_{10}(1-c_0), \ N_{psr} = 500$} \\
\hline
& Model (1) & Model (3) & Model (1) & Model (3)\\
 \hline\hline
  -1 & 0.24 & 0.16 & 0.10 & 0.03  \\
 -2 & 1.2 & 0.34 & 0.47 & 0.14  \\ 
 -3 & 11 & 3.2 & 4.43 & 1.28 \\ [1ex] 
 \hline
\end{tabular}
\caption{\it Tabulated results of Figure \ref{fig:figure8} detailing the 1-$\sigma$ resolution of each model, for $N_{psr} = 200, 500$.}
\label{table:2}
\end{table}

By
performing the same kind
of Fisher analysis as described in the previous sub-section --  always with the log-likelihood \eqref{fisher1} -- we can place constraints on the quantity $c_0$, by integrating sources up to a given redshift value. Fig \ref{fig:figure8} (and Table \ref{table:2})
 demonstrates the error bounds for $c_0$ for population model $(1)$ (first row) and $(3)$ (second row). {Model
 $(2)$ is an intermediate case among these (as can be guessed by Fig \ref{fig:figure7}) and we do not represent the
 results explicitly.}
 We  restrict the integration of sources up to $z = 1$  for both source population models, highlighting the increased precision of $1-c_0$  for
  different values of monitored pulsar number $N_{psr}$, and population model choices. Binaries at low redshift (or sources inducing large dispersion) will produce the dominant signal and amplify the magnitude of the deviation due to modified gravity. This effect appears particularly prominent when comparing across population models - population model (1) reproduces a more homogeneous population of low-mass black hole binaries distributed relatively broadly across redshift. In comparison, population model (3) is more inhomogeneous, and contains fewer, but more massive (and thus, louder) binaries distributed in a more concentrated nearby region. It would therefore be expected that this distribution provides a lower error in $1-c_0$, as the signal sources will be more clearly resolved. 
  
  The results indicate that, within the hypothesis behind our analysis,  values of $1-c_0$ of order $10^{-2}$
  can be tested around the level of 10\% accuracy by monitoring many pulsars; however, the accuracy
  rapidly degrades as we reduce the values of
   $1-c_0$. Including the population model appears to reduce the relative error by approximately one order of magnitude - which would be anticipated due to a more precise source-modelling scheme (as the results demonstrated in figure \ref{fig:figure8} are found by marginalizing over redshift, with the upper limit of the integration in $z$ appearing as a Fisher parameter). We can interpret this upper limit as the redshift to the furthest source, and characterizes the spread of the source distribution. For this model, we can not take the redshift to be too large, as at large redshift, the fit model $\Tilde{\Omega}_{\rm GW}$ begins to diverge.
   We also see from Table \ref{table:2} that for population model (3), it may be possible to observe deviations on the order of $1-c_0 = 10^{-3}$ with higher pulsar numbers. It would be very interesting  to refine our discussion, and consider more realistic modelling for
   the pulsar noise or explore further source population models, for instance (4) and (5) in \cite{Sato-Polito:2023spo} or phenomenological models as in \cite{Casey-Clyde:2021xro} to discover their ability to constrain deviations from $c_0 = 1$; further pushing the boundaries of smaller detectable deviations.

\section{Conclusions}
\label{sec_conc}

Several well motivated scenarios of modified
gravity, when applied to cosmological settings,
predict that the speed of GW is different from that of light. 
We generalized Phinney's `practical theorem' to scenarios
with modified GW dispersion relations. The generalization is particularly interesting
to consider in models predicting a frequency-dependent GW speed, which
changes in a sizeable way within  the frequency band of a given detector. Adopting a well motivated
ansatz for modified GW dispersion relations, we pointed
out that they can lead to a localized distortion on the SGWB frequency profile, 
potentially detectable by future experiments. We focused
on the SGWB produced in the initial inspiral
phase by supermassive black hole binaries 
in circular orbits. By means of a dedicated  Fisher
analysis, we forecasted opportunities   to detect modified gravity effects 
with PTA experiments,
monitoring a large number of pulsars. We pointed out that 
these effects, besides modified
gravity, also depend on the details of the source population, alongside redshift
position of the GW sources. If detected, the
effects we have highlighted can be used as cosmic ladders to infer
cosmic distances. It would be interesting to further explore
these ideas, which require careful discussion and analysis of intrinsic noise pulsar properties,  and
further
considerations of  possible 
 degeneracies with astrophysical effects.  While we focused on
 applications to pulsar timing arrays working at nano-Hertz GW
 frequencies, our generalization of Phinney's theorem may readily be applied
 to other frequency bands and to ground-based  or space-based detectors. For example, to study possible modifications of the frequency profile of the SGWB caused
by stellar origin binary black holes in the LISA band \cite{Babak:2023lro}, which might be detected for example using
the techniques proposed in \cite{Caprini:2019pxz}. 
 We hope to return to investigate these possibilities in  separate studies.

\subsection*{Acknowledgments}
We are
partially funded by STFC grant ST/X000648/1.  
For the purpose of open access, the authors have applied a Creative Commons Attribution licence to any Author Accepted Manuscript version arising.
Research Data Access Statement: No new data were generated for this manuscript.

{\small
\addcontentsline{toc}{section}{References}
\bibliographystyle{utphys}

\bibliography{phinney_gen}
}

\end{document}